\documentclass{article}

\usepackage{arxiv}

\usepackage[utf8]{inputenc} % allow utf-8 input
\usepackage[T1]{fontenc}    % use 8-bit T1 fonts
\usepackage{hyperref}       % hyperlinks
\usepackage{url}            % simple URL typesetting
\usepackage{booktabs}       % professional-quality tables
\usepackage{amsfonts}       % blackboard math symbols
\usepackage{nicefrac}       % compact symbols for 1/2, etc.
\usepackage{microtype}      % microtypography
\usepackage{lipsum}		% Can be removed after putting your text content
\usepackage{graphicx}

\usepackage{csquotes}

\title{Towards Somaesthetics Inspired Games:  Exploring the Influence of a Mirror Effect on Self-presentation in a Public Setting}

\author{{Fiona Guerin, Alice Rey, Enis Caliskan, Erik Kynast,  Andreas Zimmerer,   Ilhan Aslan, and Elisabeth Andr{\'e}}\\
Human-Centered Multimedia Lab\\
Augsburg University\\
Germany\\
\texttt{name.lastname@student.uni-augsburg.de, lastname@hcm-lab.de} 
}

\begin{document}
\maketitle

\begin{abstract}
We report on an initial user study, which explores how players of an augmented mirror game, self-style or self-present themselves when they are allowed to see themselves in the mirror compared to when they do not see themselves. To this end, we customized an open source fruit slicing game into an interactive installation for an architecture museum and conducted with 36 visitors a field study. Based on an analysis of video recordings of participants we identified, for example significant differences in how often participants smile. Ultimately, presenting a self-image to gamers in a social setting resulted in behavior change, which we argue could be utilized carefully from a Somaesthetics perspective  as an experience design feature in future games. 		
	
\end{abstract}

% keywords can be removed

\section{Introduction}
\label{sec:introduction}
With the current rise of social augmented reality games like Pokemon Go\footnote{\url{https://www.pokemongo.com/}}, more and more people play engaging games in public. Common to these games are that players interact (sometimes involuntarily) with their surrounding social context. There are usually other people (spectators) involved, which lack the information given in the game and therefore are only able to observe the visible actions and movements of the playing person. Furthermore, some games might push the playing person to subconsciously do things they normally would not do in a public setting because they might feel embarrassed doing so.
The combination of potentially doing unreasonable actions and unwanted movements might lead to players feeling insecure and quitting the game because they do not want to take the risk of feeling embarrassed in a public context.

However, most games in public space (e.g., Augmented Reality Games) currently do not address the issue of giving the players feedback on how they act in the surrounding public context. We believe that Somaesthetics could provide a stance addressing related issues. Richard Shusterman (\cite{shusterman2008body}, p. 111) describes the field of Somaesthetics broadly as \emph{``a critical study and meliorative cultivation of the soma[the living body] as a sight both of sensory appreciation (aesthesis) and creative self-fashioning''}. He
argues (simply put) that our bodies are our main tools to experience our surrounding world and that they are also tools which we use for self-presentation, self-styling, and self-fashioning purposes. Consequently both how we feel and ho we present ourselves are shaped by body-centric practices and offering (digital) opportunities to improve or reflect on our self-presentation, self-styling, and self-fashioning would be beneficial. HCI research taking inspirations from Somaesthetics has been increasing, with recent research addressing tangibles for shared experiences and social resonance \cite{aslan2020pihearts}, affective mirrors in smart homes \cite{dang2019towards},  human-drone interaction \cite{la2020drone},  and for ideation and design-driven critique (e.g., \cite{eriksson2020ethics, tennent2020soma, svanaes2020designer}). 

Inspired by Somaesthetics, we propose that games should provide user feedback on self-presentation to help them control their appearance and for example feel secure. To this end, we designed an augmented mirror game, which a person can play in front of large screen. To study  to what extent an additional mirror feedback on players' self-presentation influences their behavior in a public context  we conducted a field study in an architecture museum with 36 visitors. We found significant differences considering facial expressions and gestures. Participants smiled more move more elegantly when they saw themselves in the mirror effect.
Before, we describe our prototype, study setup, and results in detail we provide the background, including related work in the next section.

%Overall, there is indeed a minor effect on the self-presentation of people when playing games in a public context while getting feedback on their own appearance. This also leads to the conclusion that there is a certain supportive effect of displaying the image of a person for public games.

%The contribution of this work is that we provided the first empirical study on how people change their self-presentation while playing Augmented Reality Games in a public context with and without feedback on their appearance.

\begin{figure}
\center
  \includegraphics[width=0.8\textwidth]{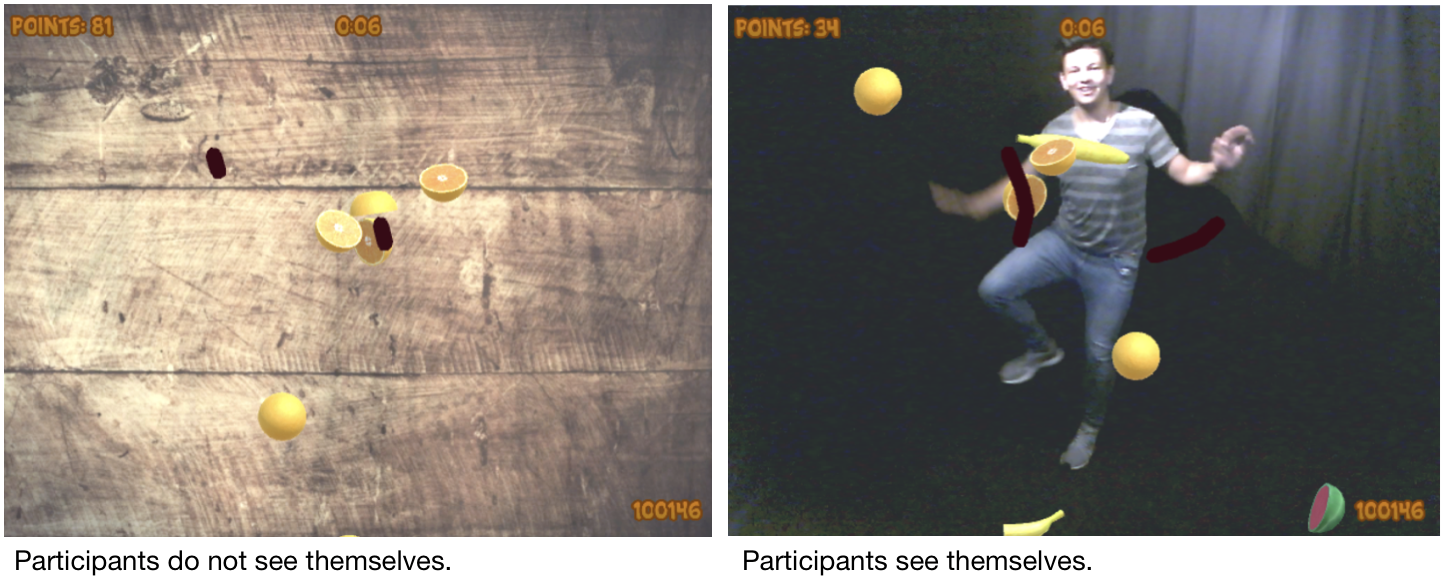}
  \caption{We compare an augmented reality game in which participants see themselves with the same game in which participants do not see themselves, and we ask to what extent their own image influences their self-presentation.}
  \label{fig:teaser}
\end{figure}

\section{Related Work}
\label{sec:relatedwork}
The work is partly inspired by research in novel interaction techniques conducted by researchers at the human-centered multimedia lab, which include 
mid-air and proxemic interaction (e.g., \cite{aslan2016design,aslan2017pre,aslan2016pen+,aslan2019put, bittner2019smarthomes})  
and intelligent  and adaptive systems (e.g., \cite{flutura2018drinkwatch,ritschel2019irony,ritschel2019personalized,rathmann2020towards}). In the following we put the core characteristics of our work in relation to previous work: First, we define Mixed Reality, Virtual Reality, and Augmented Reality. Second, we differentiate user-centered and non-user-centered augmentation. Third, we inspect Augmented Reality Mirrors. Fourth, we differentiate between single and multiple user games. Fifth, we examine Augmented Reality Mirrors for the entertainment industry. Sixth, we define why, when, and how subjects present themselves to others.

\subsection{Mixed Reality, Virtual Reality, Augmented Reality}
Mixed Reality is derived from Virtual Reality and it generalizes Augmented Reality. Mixed Reality integrates digital objects into the real world. Users of Mixed Reality simultaneously perceive reality and virtuality. Contrarily, Virtual Reality immerses its users in a virtual, computer-generated world and isolates them from reality. \cite{mixed_reality} In Mixed Reality, real elements can dominate virtual elements and vice versa. In Augmented Reality, real elements always dominate virtual elements. \cite{mixed_reality} An exemplary Mixed Reality application is a virtual game that integrates the mirror image of its players. An exemplary Augmented Reality application places name tags on real objects. An exemplary Virtual Reality application presents the 3D model of a building to its user on Virtual Reality glasses.

\subsection{User-Centered vs. Non User-Centered Augmentation}
A user-centered augmentation enhances the users' perception of themselves with virtual objects. Their behavior and movements are mapped on virtual objects and they perceive themselves from the avatar perspective. Thus, the user is transformed into a character or avatar. Studies of this transformation have shown, that users that are transformed into characters such as Mr. and Mrs. Potato or famous celebrities try to imitate the behavior of the character by changing their tone of voice or their movements. Non-centered Augmentation focuses on adding virtual content to reality rather than trying to change it. This makes the augmentation external to the user and centers on the interaction of the user with the augmented objects \cite{portales2018interacting}.

\subsection{Augmented Reality Mirrors}
Augmented Reality Mirrors add a new layer to the regular image of the scene reflected by mirrors. In this additional layer, virtual objects are added to the scene and thus augment the perceived reflection. An Augmented Reality Mirror consists of two parts: A tracking technology that perceives users or virtual objects in real-time and a display technology that merges and displays the virtual objects and real perception. Tracking technologies can be based on different kinds of sensors, but with the rise of image recognition through machine learning techniques, visual sensors such as cameras are preferred. This kind of recognition makes additional attachments to the interacting individual or object unnecessary and provides intuitive gesture interaction.

Early work on ``smart'' mirrors  in buildings has study how  messages and reminders could be integrated in mirros (e.g.,  \cite{Helal:2005:Smarthouse}). Since then many fellow researcher have explored how mirrors could be extended with new digital features  (e.g., \cite{Besserer:2016:FitMirror,Gold:2016:MirrorFramework,Njaka:2018:MirrorAuth,Yang:2018:iMirrorLights}). 

There are two kinds of output technologies for Augmented Reality Mirrors. The first kind is a complete video-based approach in which the reality and the virtual objects are both emitted via a video output. The second kind is a mixture of a semi-transparent mirror and a display behind. This way, the reality is reflected by the mirror and the virtual objects are added on a screen or projection behind the real mirror \cite{portales2018interacting}. A system can either deal with a single user or multiple users. In the multi-user scenario, the users can either perform tasks simultaneously and independently or it could be a collaborative task, whereas only one user can use the system or perform tasks in the Single User version \cite{portales2018interacting}.

\subsection{Augmented Reality Mirrors for Entertainment}
Augmented Reality Mirrors are not only for scientific purposes. They offer an application in the entertainment industry which was already realized by Playstation in 2003 with the camera EyeToy\footnote{\url{https://www.playstation.com/de-de/games/eyetoy-play-ps2/}}. In this version of EyeToy, the movements of the user were used as interaction points. The follow-up camera EyeToy Play2\footnote{\url{https://www.playstation.com/de-de/games/eyetoy-play-2-ps2/}} by Playstation in 2004 additionally visualized the camera video to screen and created an Augmented Reality Mirror. But not only Playstation used the Augmented Reality Mirrors. The Xbox 360 was extended with a Kinect\footnote{\url{https://support.xbox.com/en-US/xbox-360/accessories/kinect-sensor-setup}} sensor to allow a similar effect as the EyeToy Play2. Another entertainment purpose was implemented by SnapChat\footnote{\url{https://www.snapchat.com/}}. SnapChat is an application that allows the user to modify the perceived camera image of their smartphones with filters or augmented objects. These improvements in the entertainment industry indicate, that users are well-disposed towards the idea of augmentation in their mirrored image and seem to have a positive conception towards this technology.  \cite{portales2018interacting}
When a digital game augments reality, its players have a better experience of the game. We augment a game by depicting our reality in it and placing virtual elements on it \cite{mixed_reality_games}.

\subsection{Self presentation}
By self-presenting ourselves, we want to create, modify, or maintain a certain impression with others \cite{impression_management}.
We self-present ourselves for different reasons: First, to facilitate social interaction \cite{the_presentation_of_self_in_everyday_life}. Second, to gain material and social reward \cite{interpersonal_perception}. Third, to construct our identity \cite{impression_management}.
A teacher, for example, may act as authoritarian, firstly to calm their pupils during class, secondly to keep their job, and thirdly to consider themselves dignified and thus to accept themselves. 
We self-present ourselves on different occasions: First, when foreign judgment influences external reward \cite{drama_and_the_self_in_social_interaction, impression_management}. Second, when others are aware of us \cite{aspects_of_self_and_the_control_of_behavior}. Third, when others ignore us \cite{self_consciousness_and_social_anxiety}. Fourth, when a situation is not familiar to us \cite{self_presentation_in_everyday_interactions}. 

We therefore pay special attention to our self-presentation, for example when we are interviewed for a job and when we give a presentation.
When we self-present ourselves, we select our words, for example, we boast \cite{self_consciousness_and_self_presentation}. We pay attention to our appearance, for example, we pay attention to hair, figure, and clothing \cite{self_consciousness_and_self_presentation}. And we regulate our movements and thus we regulate facial expressions, gestures, gait, and posture \cite{toward_an_ecological_theory_of_social_perception}.

\section{Augmented Reality Game}
\label{sec:augmentedrealitygame}
\subsection{Goal}
We investigate how one's image influences self-presentation in an Augmented Reality Game. We ask how participants change their facial expressions and their gestures when they see themselves in the Augmented Reality Game compared to when not. 

For this, we create an Augmented Reality Game. The participants should feel like standing in front of a mirror while interacting with the Augmented Reality presented to them in the mirror. We choose an Augmented Reality Arcade Game where the goal is to slice fruit that flies into the screen from underneath. The feet and hands are the interaction media of our Augmented Reality Game. 

\subsection{Design Choices}
Based on the presented related work, the following design choices were made.

\subsubsection{Non User-Centered Augmentation}
This paper focuses on the self-presentation of users and it would, therefore, be disadvantageous to use the user-centered approach. It could not be excluded that the user changes their behavior to fit into the perception of the augmented character. To observe how the mirror image influences one's self-presentation in the Augmented Reality Game, we use the non-centered approach.

\subsubsection{User recognition}
For this paper, a camera-based solution was implemented to analyze user interaction. This is based on the independence of additional equipment, such as markers or other sensors, which could distract or influence the user.

\subsubsection{Video-based mirror approach}
A video-based mirror setup simplifies the implementation concerning the mapping of the reality and the augmented objects. Furthermore, the video-based setup is portable and only requires a screen whereas the semi-transparent mirror approach depends on an additional see-through mirror. For the sake of simplicity and cost, we chose the video-based mirror.

\subsubsection{Single User}
To focus on the self-image of one user, the Single User approach was preferred to cancel out possible distractions and the sense of competition originating through another user.

\subsubsection{Entertainment factor}
The examples for Augmented Reality Mirrors used by the entertainment industry indicate a positive reception of this technology. Based on this assumption, we chose an entertaining game setup.

\subsection{Mirror Setup}
We decided to not use an actual mirror for our study but a big TV screen together with a webcam at the center on top of the screen. On the TV we display the webcam video stream to create the illusion of looking in a mirror and add the fruits as a second layer on top of the webcam video. To slice the fruits, the participants have to hover with the left or right hand over them in the mirror image. To enable the slicing the participants have to stand on one foot. Otherwise, nothing is happening when their hand hovers a fruit in the mirror image. The software needs two webcam video streams. One for mirror illusion and one for the Pose Recognition. To make sure the mapping of the coordinates received from the pose recognition is the same as the ones of the mirror background, we use one camera only and split the input into two webcam video streams with SplitCam.

\subsection{Gesture and Pose Recognition}
For being able to use the hands and feet as interaction medium we are using a real-time system that can detect human body keypoints. We use the open-source software OpenPose. OpenPose can detect those key points in 2D. We used as an input for the OpenPose Tool the data that we received from the webcam. OpenPose detects gesture key points based on the video stream as input. OpenPose creates an output stream to the standard output containing the key points in the BODY\_25 format. This format contains the coordinates of 25 body parts as well as the probability of body parts being at the given coordinates. We check the probability of the feet and hand before we use the coordinates for our computation to make sure these are sufficiently correct. In our case, we chose 50\% as an acceptable limit. The standard output is fetched by our software and analyzed with string operations. To improve the performance of the gesture recognition we restricted the number of people that OpenPose recognizes to 1 and reduced the net-resolution.

\subsection{Game Implementation}
As a starting point, we used the Open Source Software fruit-ninja-replica implemented in C\# with Unity. To adapt the software to our needs we replaced the background of the game with the video stream captured from the webcam. The slicing functionality is implemented in fruit-ninja-replica with the mouse as an interaction medium. When the mouse is clicked, slicing is enabled. By moving the mouse over fruits, they are being sliced. We replaced the mouse position with the coordinates of the left and right hand which we receive from OpenPose. This means that in our case the participants have to use their hands to slice fruits instead of the mouse. To enable the slicing the participants have to stand on one foot instead of previously keeping the mouse button clicked. We recognize this position by checking the distance of the y-coordinates of both feet.

\subsection{Fruit Slicing}
For our game, we use watermelons, bananas, and oranges as fruits. We used two assets from the Unity Asset Store. The Free Citrus Fruit Pack Asset for the orange and banana and the RPG Food \& Drinks Pack Asset for the watermelon. A fruit is marked as sliced when the position of the left or right hand and the fruit position are equal and the slicing is enabled. As soon as a fruit is marked as sliced it is replaced by two halves of the same fruit which are placed on top of each other. This creates the illusion of sliced fruit as soon as they fall apart. A collider trigger imitates the collision of the fruit and the separation of two fruit halves.

\subsection{Course Of the Game}
The augmented reality game lasts for one minute during which the participants need to slice as many fruits as possible. For every sliced fruit, they receive one point. The game differentiates between two modes. In the first 30 seconds, we imitate a mirror where the participants feel like standing in front of a mirror. They see themselves mirror-inverted and the fruits flying into the mirror from underneath over the whole screen which can be seen in figure \ref{fig:mirror_mode}. By standing on one foot and moving the arms over the fruits the participants can slice the fruits. 
We differentiate between those two modes for our study to be able to compare the behavior of the participants with and without seeing themselves. The order of the modes changes with every game, to ensure that the ordering of the modes does not influence the results. During the game, we display the overall points the participants received so far in the top left corner, the time left for each mode at the center of the top and the game ID to be able to map the game to the survey afterward in the bottom right corner.

\section{User Study}
\label{sec:userstudy}
We investigate to what extent we present ourselves differently in an Augmented Reality game when we see ourselves in it. On the one hand, we observe to what extent subjects change their facial expressions and gestures when they see themselves in the game. On the other hand, we ask the subjects to what extent they have perceived a change in facial expressions and gestures. 

When subjects present themselves in our Augmented Reality game, they present themselves in front of other people. For this purpose, we let our subjects play the Augmented Reality game in a museum. While one subject is playing, the other people stand in line for the Augmented Reality game and watch the subject. There are therefore three occasions for the subjects to present themselves in front of others: First of all, the other people are attentive to the subjects while queuing for the Augmented Reality game. Second, the subjects can expect social recognition from its audience. Third, the subjects play the Augmented Reality game in front of strangers. 

\subsection{Participants}
Our study comprises 36 participants. Because the participants are summer festival guests in an architecture museum, their gender and age are randomly distributed. No participant is involved in our research and no participant has played our game before.

\subsection{Experimental procedure}
Our hardware comprises a computer, a monitor, and one camera. Our computer contains a high-performance graphics card Nvidia GTX 1060 with 6GB RAM. Our monitor is a LG 55LM660S-ZA TV. Our camera is a Logitech Webcam C250.

We set up our lab so that our players can see themselves in our Augmented Reality game. The Augmented Reality game is started on the computer together with the gesture recognition software. The camera connected to the computer provides the video input with 30 frames per second and a hardware resolution of 640x480 pixels. The computer is then connected to the monitor. Each round will be recorded for research purposes and is not part of the actual game.

When we set up our laboratory, we pay attention to the distance between our Augmented Reality game and its participants: For our camera to recognize the participants, there has to be a distance of about 2 meters.

We conduct our study in the following seven steps:
    (1) We welcome every subject in front of our lab.
    (2) The subjects enter our laboratory one by one.
    (3) The subjects fill out our privacy policy.
    (4) We explain our Augmented Reality game to the subject. The goal of the game is to cut as many fruits as possible in a certain time. The subjects can cut fruit by moving their arms over the fruit in the Augmented Reality game. The subjects can only cut the fruit when standing on one leg.
    (5) The subjects play both rounds of our Augmented Reality game: In one round, the subjects see their mirror image in the game and in the other round, they do not see their mirror image. Half of all players first play the round with the mirror-image and then the round without their mirror-image. The other half does not see their mirror-image at first and sees their mirror-image in the second round. We film the subjects while they play the two rounds.
    (6) The subjects fill out a questionnaire in which they evaluate their self-presentation in our Augmented Reality game.
    (7) We thank the subjects and say goodbye.

\subsection{Data collection}
Our study collects self-representation data. The self-representation data describes on the one hand how we have observed the self-representation of a subject from the outside, and on the other hand how the subject perceives its self-representation. 

We evaluate the subjects' self-presentation according to how often the subjects smile and how elegantly they move their arms. While the subjects cut fruits in our Augmented Reality game, they can move their arms discreetly, flowingly, and randomly: 
Discreet movements are short and defined, thus they look like karate movements: They are elegant.
Flowing movements are long and defined, thus they look like swimming or ballet movements: They are elegant.
Random movements do not follow a defined structure: They are less elegant. 

\begin{figure}[h]
    \centering
    \includegraphics[width=0.5\linewidth,trim={0 0 0 0},clip]{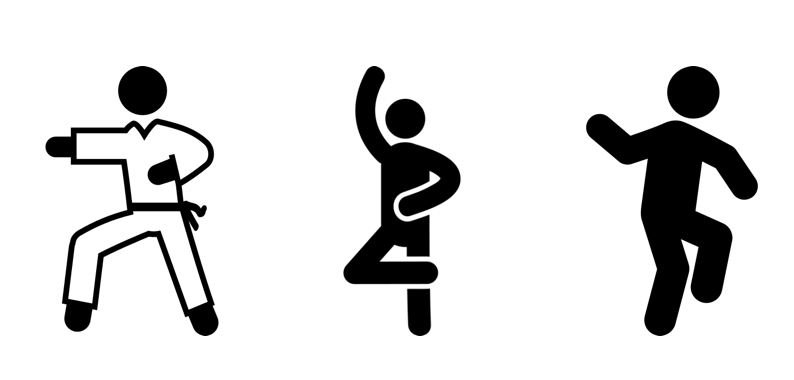}
    \caption{Discreete, flowing, and random arm movement.}
        \label{fig:data_collection:movement}
    \vspace{0mm}
\end{figure}

To observe the self-representation of subjects from the outside, we film the subjects and annotate the recordings. On the one hand, we count how often the subjects smile when they do not see themselves in the Augmented Reality game, and how often subjects smile when they see themselves in the Augmented Reality game within the same period. On the other hand, we assign the subjects' movements to one of the categories discrete, flowing, and random when they do not see themselves in the Augmented Reality game, and we assign the subject's movements to one of these categories when they see themselves in the Augmented Reality game. 

The subjects evaluate their perceived self-representation in a questionnaire. In it, they compare their facial expressions and elegance in the Augmented Reality game with and without their image. 

\subsection{Experimental design}
We design the study as an A/B test with repeated measurements. In it, we compare the self-presentations in an Augmented Reality game without one's mirror image and the self-presentations in an Augmented Reality game with one's mirror image. We examine how two factors -- the Augmented Reality game without image and the Augmented Reality game with a mirror-image -- influence four dependent variables -- the perceived facial expression, the perceived elegance of arm movement, the observed facial expression and the observed elegance of arm movement.

\section{Results}
\label{sec:results}
\subsection{Observed self-presentation}
\subsubsection{Facial expression}
We suppose that subjects smile more often when they see their image in the Augmented Reality game than when they do not see their image in the Augmented Reality game. We formulate the following alternative hypothesis:  

H1: In an Augmented Reality game, subjects smile more often when they see their mirror-image than when not. 

We calculate the average and standard deviation of the number of smiles in an Augmented Reality game without their mirror-image as well as average and standard deviation of the number of smiles in an Augmented Reality game with their mirror-image. Subjects who do not see their image in the Augmented Reality game smile 1.9 times on average and the standard deviation is 1.7. If subjects see themselves in the Augmented Reality game, they smile 2.6 times on average and the standard deviation is 1.8.

\begin{figure}[h]
    \centering
    \includegraphics[width=0.5\linewidth,trim={0 0 0 0},clip]{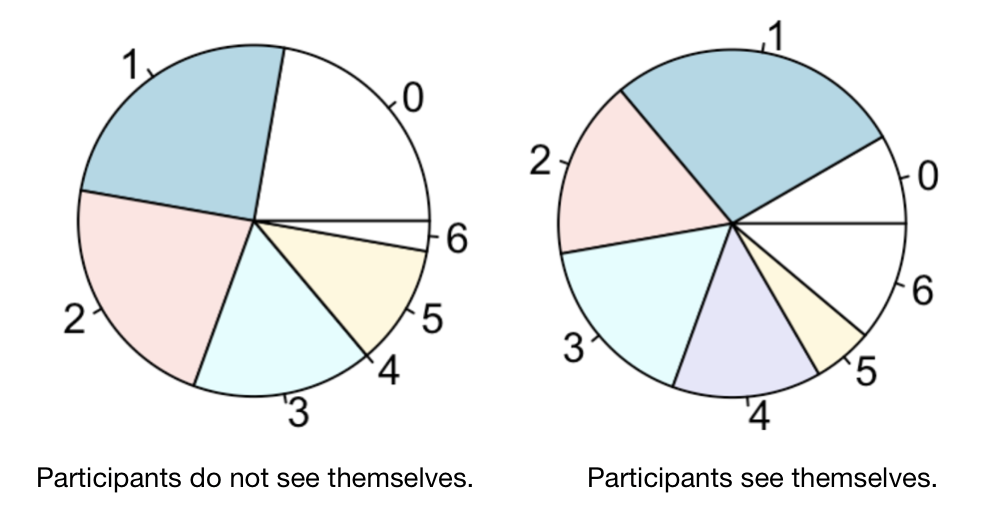}
    \caption{Number of smiles without and with image.
        \label{fig:results:observe_smiles_pie}}
    \vspace{0mm}
\end{figure}

A  t-test  comparing the scores for both conditions (i.e., with and without the mirror effect) showed that the existence of the own image has a significant effect on the number of smiles (t = 2.0, p = 0.024). Participants smiled significantly more when they saw themselves playing the game. 

\subsubsection{Gesture}
We suppose that subjects move their arms more elegantly when they see their image in the Augmented Reality game than when they do not see their image in the Augmented Reality game. We formulate the following alternative hypothesis: 

H1: In an Augmented Reality game, subjects move their arms more elegantly when they see their mirror-image than when not. 

When subjects see their mirror-image in the Augmented Reality game, they move their arms more elegantly. We calculate how elegantly subjects move their arms on a scale of 0 to 1. A value of 0 indicates a completely random movement and a 1 indicates a completely elegant movement, i.e. a discreet movement or a flowing movement. On average, the subjects move their arms rather randomly -- the scale value is 0.3 -- and the standard deviation is 0.5 if they do not see their image in the Augmented Reality game. When the subjects see their image in the game, they move their arms rather elegantly. The scale value is 0.6 and the standard deviation is 0.5. 

\begin{figure}[h]
    \centering
    \includegraphics[width=0.5\linewidth,trim={0 0 0 0},clip]{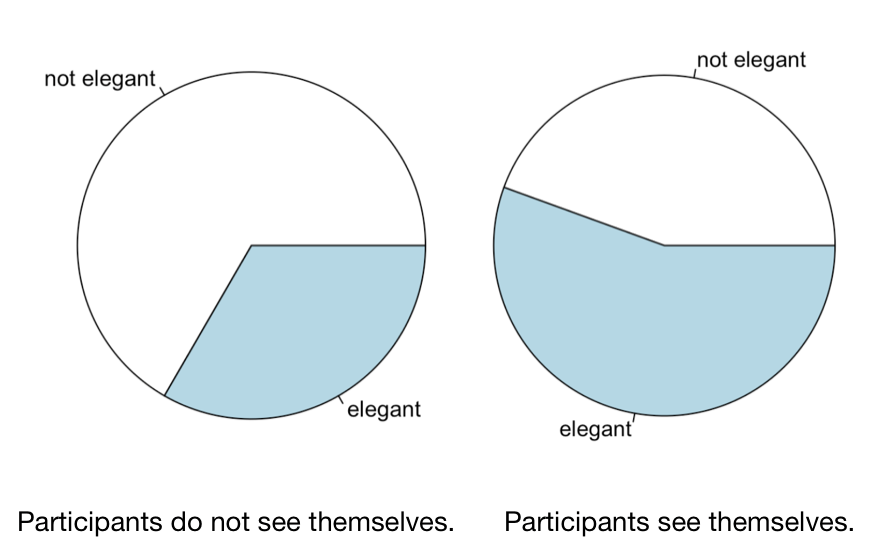}
    \caption{Elegant movements without and with image.
        \label{fig:results:observe_elegance_pie}}
    \vspace{0mm}
\end{figure}

A t-test (t = 1.7,  p = 0.052) shows that the effect of existence of the own image on elegance of movement was non-significant. However, since the p-value is slightly above the standard significance level of 5\% we tend to consider there to be a ``trend`` towards participants moving more elegant when they see themselves during game play.

\subsection{Perceived self-presentation}
In addition to collecting ``objective'' measurements considering self-presentation (i.e., video annotations) we collected subjective measurements, considering perceived difference in self-presentation. Moreover we had asked if participants agree (on a 5-point Likert scale) that they self-presented themselves when they saw themselves in the ``mirror'' compared to the condition without the ``mirror'' effect.  A provided score of 1 meant that participants did  not agree at all  with the statement and a value of 5 meant that they strongly agreed with that statement. We collected agreement scores for two statements, one addressing differences in facial expression and the other on body posture/gesture. 

Results show that participants believe to have neither changed their facial expression or their posture/gesture under both conditions (i.e., with and without the mirror effect).  For facial expression they reported a mean score of 2.7 (SD=1.1) and for body posture/gesture a mean score of 2.8 (SD=0.9). Consequently, participants seem to be mostly unaware that they ``self-style'' themselves when they see their own image.

\section{Discussion}
\label{sec:discussion}

In the beginning, we have argued that  self-presenting, self-styling, or self-fashioning is an habit we, as humans do when we interact with others, and we do this possibly even when interacting with ourselves through a ``medium'' such as a mirror or a technology utilizing a camera. For example, people undoubtedly  tend to smile and show their best side to a camera and potentially to a mirror.  Thus, self-presenting is something we often do in an ``automated'' manner. Most of us may have encounter a situation in which someone is told how beautiful or natural the picture of them is, because they seem in the picture unaware that they are being captured in that moment. Because of unawareness people possibly don't self-present or over-style themselves and the result seems special, because in truth we may self-present continuously in our interactions.   

In this paper, we presented an initial study into the notion of self-presenting or self-styling and how related behaviors are triggered in public game setting by a mirror ``effect''.  Our intention is bringing a Somaesthetics perspective towards the design of public and  Augmented Reality games. This perspective propagates that self-presenting is inherent to having and acting through a living body, which for example also reacts to external stimuli (e.g. when we freeze) in a automated manner. Thus, we don't consider acts of self-presentation as something negative that people should feel shame for  but something potentially aesthetic and positive in how we use our bodies (e.g, ``strike poses'' and smile). Ultimately, we believe that self-presenting is something inherent to how we interact in social settings and that it is worth analyzing effects of self-presenting, enabling opportunities for self-presentation, and 
potentially bringing them to the foreground for users and designers to reflect on and design with and for it.

Our results show that there is a significant difference in how much participants smile and (a trend for) how beautiful they move. When we asked participants if they thought they had acted differently, participants seem to be mostly unaware that they ``self-present'' themselves differently (and more) when they see their own image. Of course there is a chance that participants do not want to admit  self-presenting themselves. Richard Shusterman discusses in his books (e.g., \cite{shusterman2008body}) how the body is often culturally perceived as  something ``sinful'' and admiring ones own body is usually something negative. But in reality and contrary to the notion that physical appearance should not matter, people spend a lot of time optimizing their appearance, including practices such as bodybuilding, putting on makeup, or beauty surgeries.  Not all practices are healthy and some people (e.g., with an unhealthy tendency to narcissism) may be more vulnerable to new technology designs which enable new forms of self-presentation. We hope our research will inspired and contribute to carefully analyzing and designing self-presentation opportunities enabled by technology and how we can improve future designs, which we believe is more important than ever in an age of potentially harmful altered self-images on social media (e.g., \cite {lowe2018self}).

\subsection{Limitations and Future Work}

Our Augmented Reality game has been unexpectedly challenging for many players. Since we conducted the study in an architecture museum many of the participants were older. Compared to an easier game, the players may not have realized as much that they had self-represented themselves in it. Challenges in our game are physical, mental, and technical challenges:
    (1) Physical challenge: A player always has to stand on one leg. 
    (2) Mental challenge: No player has played our game before.
    (3) Technical challenge: Many players are not familiar with Augmented Reality. 

There are two main areas we would encourage future research to explore.
The first one is adding a multiplayer component to the game to test whether the social connections change the behavior of the players. Good questions to ask would be, whether the real-life social structures influence the behavior of players. Does the focus of players change based on social connection to the other players? The second one is exploring different levels of cognitive load and how it may influence participants self-presenting behavior. Based on our observations and our questionnaire, participants' cognitive resources decreased when they saw their mirror-image. It seems that self-presentation opportunities result in participants requiring more cognitive resources since they tend to use some of their resources willingly or unaware for self-presenting.  An important limitation of our study is that our participants were older and therefore it would be interesting how younger participant self-present and if the physical demand (for example standing on one foot) may effect their self-presenting behavior.

\section{Conclusion}
\label{sec:conclusion}
In this paper, we reported on user study in an architecture museum with 36 participants. We explored how participants change their behavior (i.e., self-presenting) when they can see their mirror-image while they play a game in a public setting.  

We found based on analyzing video recordings of the players that when players see their mirror-image in a public game, they self-present themselves more, i.e. they smile significantly more and they move more ``elegantly''.  
We have motivated and discussed why self-presentation  opportunities should be considered as both a matter worth analyzing and design material for public games (especially augmented reality games). 
We hope that our research will contribute to future games that will carefully integrate self-presentation opportunities, considering them as  necessary and benefitial for us as embodied and social beings and that carefully allowing for self-presentation might improve our (psychological) wellbeing and ultimately improve player experiences in public settings.

\bibliographystyle{unsrt}
\bibliography{references}

\end{document}